\documentstyle[art10]{article}
\begin{document}

\begin{titlepage}

\title{Octonion X,Y-Product $G_{2}$ Variants}

\author{Geoffrey Dixon \thanks {* Happy Birthday to Larry
Horwitz.  If there's a next time you must come.} \\ 
Department of 
Physics \\ Brandeis University \\ Waltham, MA 02254 \\
email: dixon@binah.cc.brandeis.edu}

\maketitle

\begin{abstract} 
The automorphism group $G_{2}$ of the octonions changes when
octonion X,Y-product variants are used.  I present here a
general solution for how to go from $G_{2}$ to its
X,Y-product variant.
\end{abstract}

\end{titlepage}

\section*{1. X,Y-Product.}

Let $e_{a}, \; a=0,...,7$, be a basis for the octonion algebra, 
{\bf O}.  There are 7680 distinct ways to define an octonionic
multiplication on these 8 symbols such that for all $a,b \in
\{0,...,7\}$ there exists some $c \in \{0,...,7\}$ satisfying
\begin{equation} 
e_{a}e_{b} = \pm e_{c}.
\end{equation} 
My favorite is to let $e_{0} = 1$ be the identity, and let
\begin{equation} 
e_{a}e_{a+1} = e_{a-2}, \; \; a \in \{1,...,7\},
\end{equation} 
and $a+1, a-2 \in  \{1,...,7\}$ are computed modulo 7.  See
{\bf [1 - 4]}.

Whatever your starting multiplication, it's still
true that for all unit elements $X,Y \in {\bf O}$, and general 
$A,B \in {\bf O}$, if we replace the {\bf O} product, $AB$,
with
\begin{equation} 
A\circ_{X,Y}B = (AX)(Y^{\dagger}B),
\end{equation} 
the resulting algebra is again a copy of the octonions, which
I denote ${\bf O}_{X,Y}$.

In the special case, $X=Y$, we denote the resulting copy of
the octonions, ${\bf O}_{X}$ {\bf [5]}.  This product
satisfies 
\begin{equation} 
A\circ_{X}B = (AX)(X^{\dagger}B) = X((X^{\dagger}A)B) =
(A(BX))X^{ \dagger }.
\end{equation} 

\section*{2. Roots.}

Any unit octonion can be expressed as
\begin{equation} 
X = exp(\theta x) = \cos \theta + x \sin \theta,
\end{equation} 
where $x$ is a unit octonion with no real part (so in (5) $x$
plays a part similar to the complex unit $i$).  If $\theta \ne
0$ or $\pi$ the n$^{th}$ roots of $X$ are easily defined:
\begin{equation} 
X^{\frac{1}{n}} = exp [(\frac{\theta + 2\pi k}{n})x], \; \;
k=0,...,n-1.
\end{equation} 
Although the n$^{th}$ roots of $X$ are not uniquely defined,
it is clear from (6) that as long as $X \ne \pm 1$, then $X$
and all its n$^{th}$ roots commute.  Consequently any product
of three octonions, two of which are n$^{th}$ roots of 
 $X$, must associate.  (If $X=-1$, for example, then the
square roots of $X=-1$ are all the elements of the 6-sphere of
unit octonions with no real part.)

\section*{3. $G_{2X}$.}

Therefore, given a unit octonion $X \ne \pm 1$, a given cube
root, arbitrary $A,B \in {\bf O}$, and using (4),
\begin{equation} 
\begin{array}{rl}
A\circ_{X}B & = (AX)(X^{ \dagger }B) \\ \\
& = (AX^{2/3}X^{1/3})(X^{-1/3}X^{-2/3}B) \\ \\
& = X^{1/3}[(X^{-1/3}AX^{2/3})(X^{-2/3}B)] \\ \\
& = X^{1/3}[(X^{-1/3}AX^{1/3}X^{1/3})(X^{-1/3}X^{-1/3}B)] \\ 
\\
& = 
X^{1/3}\{[(X^{-1/3}AX^{1/3})(X^{-1/3}BX^{1/3})]X^{-1/3}\} \\
\\ 
& = 
X^{1/3}[(X^{-1/3}AX^{1/3})(X^{-1/3}BX^{1/3})]X^{-1/3}. \\
\end{array} 
\end{equation} 

Note that if $X= \pm 1$, then $A\circ_{X}B = AB$, in which
case (7) implies
\begin{equation} 
X^{-1/3}(AB)X^{1/3} = (X^{-1/3}AX^{1/3})(X^{-1/3}BX^{1/3}).
\end{equation} 
That is, if $U$ is a $6^{th}$ root of unity, then the action
$A \rightarrow UAU^{ \dagger }$ is an element of $G_{2}$, the
automorphism group of {\bf O}.

From now on let $G_{2}$ denote the automorphism group of our
original copy of {\bf O}, and $G_{2X,Y}$ and $G_{2X}$ the
automorphism groups of ${\bf O}_{X,Y}$ and ${\bf O}_{X}$ (all
copies od {\bf O} are isomorphic, of course).  Let $LG_{2}$,
$LG_{2X,Y}$ and
$LG_{2X}$ be their respective Lie algebras.

For any $A \in {\bf O}$, define $A_{L}$ and $A_{R}$ mapping
${\bf O} \rightarrow {\bf O}$ by
\begin{equation} 
A_{L}[B] = AB, \; \; \; A_{R}[B] = BA.
\end{equation} 

Let $g \in LG_{2}$, which satisfies
$$
g[AB] = (g[A])B + A(g[B]).
$$
Therefore, if $X \ne \pm 1$ is a unit octonion, and $X^{1/3}$
a chosen cube root, then \newpage
\begin{equation} 
\begin{array}{l}
X^{-1/3}_{R}X^{1/3}_{L}gX^{-1/3}_{L}X^{1/3}_{R}[A\circ_{X}B] = 
X^{-1/3}_{R}X^{1/3}_{L}g[X^{-1/3}(A\circ_{X}B)X^{1/3}] \\ \\
=
X^{-1/3}_{R}X^{1/3}_{L}g[(X^{-1/3}AX^{1/3})
(X^{-1/3}BX^{1/3})]
\\ \\
= X^{-1/3}_{R}X^{1/3}_{L}
[g[X^{-1/3}AX^{1/3}](X^{-1/3}BX^{1/3}) + 
(X^{-1/3}AX^{1/3})g[X^{-1/3}BX^{1/3}]] \\ \\

= X^{-1/3}_{R}X^{1/3}_{L}
[gX^{-1/3}_{L}X^{1/3}_{R}[A](X^{-1/3}BX^{1/3}) + 
(X^{-1/3}AX^{1/3})gX^{-1/3}_{L}X^{1/3}_{R}[B]]
\\ \\

= X^{-1/3}_{R}X^{1/3}_{L}
[(X^{-1/3}(X^{-1/3}_{R}X^{1/3}_{L}gX^{-1/3}_{L}X^{1/3}_{R}
[A])X^{1/3})
(X^{-1/3}BX^{1/3}) \\
+ 
(X^{-1/3}AX^{1/3})
(X^{-1/3}(X^{-1/3}_{R}X^{1/3}_{L}gX^{-1/3}_{L}X^{1/3}_{R}
[B])X^{1/3})]
\\ \\

= 
(X^{-1/3}_{R}X^{1/3}_{L}gX^{-1/3}_{L}X^{1/3}_{R}
[A])\circ_{X} (B) +  A\circ_{X}
(X^{-1/3}_{R}X^{1/3}_{L}gX^{-1/3}_{L}X^{1/3}_{R}
[B]). \\

\end{array} 
\end{equation} 

Therefore,
\begin{equation} 
\begin{array}{l}
LG_{2X} =
X^{-1/3}_{R}X^{1/3}_{L}LG_{2}X^{-1/3}_{L}X^{1/3}_{R} \\ \\
G_{2X} = X^{-1/3}_{R}X^{1/3}_{L}G_{2}X^{-1/3}_{L}X^{1/3}_{R}.
\\ \\
\end{array} 
\end{equation} 

\section*{4. $G_{2(1,Z)}$.}

The next step to a completely general solution to $G_{2(X,Y)}$
is the case $G_{2(1,Z)}$, $Z \ne \pm 1$, with the altered
product
\begin{equation} 
\begin{array}{rl}
A\circ_{1,Z}B & = A(Z^{ \dagger }B) \\ \\
& = (AZ^{-2/3}Z^{2/3})(Z^{-2/3}Z^{-1/3}B) \\ \\
& = Z^{2/3}((Z^{-2/3}AZ^{-2/3})(Z^{-1/3}B)) \\ \\
& = Z^{2/3}((Z^{-2/3}AZ^{-1/3}Z^{-1/3})(Z^{1/3}Z^{-2/3}B)) \\
\\
& =
Z^{2/3}((Z^{-2/3}AZ^{-1/3})(Z^{-2/3}BZ^{-1/3}))Z^{1/3}
\\ \\
\end{array} 
\end{equation} 
Therefore, in much the same way as was done above we prove
\begin{equation} 
\begin{array}{l}
LG_{2(1,Z)} =
Z_{R}^{1/3}Z_{L}^{2/3}LG_{2}Z_{L}^{-2/3}Z_{R}^{-1/3}, \\ \\
G_{2(1,Z)} =
Z_{R}^{1/3}Z_{L}^{2/3}G_{2}Z_{L}^{-2/3}Z_{R}^{-1/3}. \\
\end{array} 
\end{equation}

\section*{5. $G_{2(X,Y)}$.}

Finally, as I have noted elsewhere {\bf [4]},
\begin{equation} 
A\circ_{X,Y}B = (AX)(Y^{ \dagger }B) 
= A\circ_{X}(Z^{ \dagger }\circ_{X}B),
\end{equation} 
where $Z=YX^{ \dagger }$ is the identity of the X,Y-product. 
Anyway, using the results above,
\begin{equation} 
\begin{array}{l}
(AX)(Y^{ \dagger }B)  =
A\circ_{X}(Z^{ \dagger }\circ_{X}B) \\ \\
=Z^{2/3}\circ_{X}((Z^{-2/3}\circ_{X}A\circ_{X}Z^{-1/3})
\circ_{X}
(Z^{-2/3}\circ_{X}B\circ_{X}Z^{-1/3}))\circ_{X}Z^{1/3} \\ \\
=X^{1/3}\{(X^{-1/3}Z^{2/3}X^{1/3})
[<(X^{-1/3}Z^{-2/3}X^{1/3})(X^{-1/3}AX^{1/3})
(X^{-1/3}Z^{-1/3}X^{1/3})>\bullet \\
<(X^{-1/3}Z^{-2/3}X^{1/3})(X^{-1/3}BX^{1/3})
(X^{-1/3}Z^{-1/3}X^{1/3})>](X^{-1/3}Z^{1/3}X^{1/3})\}X^{-1/3},
\\
\end{array} 
\end{equation}
implying in the way that (11) and (13) were derived that
\begin{equation} 
\begin{array}{l}
LG_{2X,Y} = \\
X_{R}^{-1/3}X_{L}^{1/3}(X^{-1/3}Z^{1/3}X^{1/3})_{R}
(X^{-1/3}Z^{2/3}X^{1/3})_{L}\bullet \\ LG_{2}\bullet \\
(X^{-1/3}Z^{-2/3}X^{1/3})_{L}(X^{-1/3}Z^{-1/3}X^{1/3})_{R}
X_{L}^{-1/3}X_{R}^{1/3}, \\ \\
G_{2X,Y} = \\
X_{R}^{-1/3}X_{L}^{1/3}(X^{-1/3}Z^{1/3}X^{1/3})_{R}
(X^{-1/3}Z^{2/3}X^{1/3})_{L}\bullet \\ G_{2}\bullet \\
(X^{-1/3}Z^{-2/3}X^{1/3})_{L}(X^{-1/3}Z^{-1/3}X^{1/3})_{R}
X_{L}^{-1/3}X_{R}^{1/3}. \\
\end{array} 
\end{equation}
Note that if $X=Y$, so $Z=1$, then (16) reduces to (11) if we
choose $Z^{2/3}=1$, and if $X=1$, and we choose $X^{1/3}=1$,
then (16) reduces to (13).  However, different roots of unity
could be chosen leading to variations of (11) and (13).

The result (16) is in no way unique.  I have
found at least one very different looking solution, but none
with the same symmetric appearance as (16).

\section*{6. Conclusion.}

As usual, I would emphasize that my motivation for pursuing
this arcane line of research is to gain an insight into the
workings of the mathematics, and ultimately to use that
insight to light a path to its further relationship to physics
{\bf [1]}.

\end{document}